\documentclass[twoside]{class-definitions}
\usepackage[latin1]{inputenc}
\usepackage{wrapfig,rotating}
\usepackage{amssymb,amsmath,array}

\newcommand{\dream} {{\sc dream }}
\newcommand{\C}    {\^{C}erenkov }

\begin{document}
\title{Dual-readout Calorimetry} 
\author{N. Akchurin$^1$, F. Bedeschi$^2$, A. Cardini$^3$, M. Cascella$^4$, F. Cei$^5$,
D. De Pedis$^6$, \\ S. Fracchia$^7$, S. Franchino$^8$,
M. Fraternali$^7$ , G. Gaudio$^7$, P. Genova$^9$ ,  J. Hauptman$^{10}$, \\
L. La Rotonda$^{11}$, S. Lee$^1$, M. Livan$^7$ , E. Meoni$^{12}$,  A. Moggi$^2$, D. Pinci$^6$, \\
A. Policicchio$^{11}$, J.G. Saraiva$^{13}$,  A. Sill$^1$, T. Venturelli$^{11}$,
R. Wigmans$^1$
\thanks{This work is supported by the US Department of Energy and by INFN, Italy.}
\vspace{.3cm}\\
1 - {\it Texas Tech University, Lubbock (TX), USA} \\
2 - {\it INFN Sezione di Pisa, Italy} \\
3 - {\it INFN Sezione di Cagliari, Monserrato (CA), Italy} \\
4 - {\it Dipartimento di Fisica, Universit`a di Salento, and INFN Sezione di Lecce, Italy} \\
5 - {\it Dipartimento di Fisica, Universit`a di Pisa, and INFN Sezione di Pisa Italy}  \\
6 - {\it INFN Sezione di Roma, Italy} \\
7 - {\it INFN Sezione di Pavia, Italy} \\
8 - {\it CERN, Gen`eve, Switzerland} \\
9 - {\it INFN Sezione di Pavia and Dipartimento di Fisica, Universit`a di Pavia, Italy} \\
10 - {\it Iowa State University, Ames (IA), USA} \\
11 - {\it Dipartimento di Fisica, Universit`a della Calabria, and INFN Cosenza, Italy} \\
12 -  {\it Tufts University, Medford (MA)} \\
13 - {\it LIP, Lisbon, Portugal}
}

\maketitle



\bigskip \bigskip \centerline{\bf Abstract} \bigskip  

\begin{abstract}
 The RD52 Project at CERN is a pure instrumentation experiment whose goal is to understand the fundamental limitations to  hadronic energy resolution, and other aspects of energy measurement, in high energy calorimeters. We have found that dual-readout calorimetry provides heretofore unprecedented  information event-by-event for energy resolution, linearity of response, ease  and robustness of calibration, fidelity of data, and particle identification, including energy lost to binding energy in nuclear break-up.  We believe that  hadronic energy resolutions of $\sigma/E \approx 1-2\%$ are within reach for dual-readout calorimeters, enabling for the first time comparable measurement precisions on electrons, photons, muons, and quarks (jets).    We briefly describe our current progress and near-term future plans.  Complete information on all aspects of our work is available at the RD52 website {\tt  http://highenergy.phys.ttu.edu/dream/}.
\end{abstract}

\newpage
\section{Introduction}   

 The DREAM collaboration is  an official CERN project, RD52, taking beam test data in the North Area.  Our emphasis is in dual-readout calorimetry in which each hadronic shower is measured in two nearly independent ways.  In our fiber calorimeters, the absorber is impregnated with two independent particle measuring media, scintillating fibers for all charged particles of the shower, and clear or \C fibers for the relativistic particles which are predominately the electrons and positrons from the $\gamma$-initiated showers from $\pi^0$ decay.  
 
 In our crystal calorimeters, both scintillation light and \C light are generated in the same optical volume, and the necessary separation of the two kinds of light is accomplished by using those features that distinguish \C light from scintillation light, {\it viz.}, time structure, direction, wavelength spectrum, and polarization.  We have succeeded in crystal dual-readout in all crystals we have tested:  pure $PbWO_4$,  Molybdenum-doped $PbWO_4$:$Mo$, Praseodymium-doped $PbWO_4$:$Pr$, bismuth germanate $BGO$, and bismuth silicate $BSO$.  These crystals display a wide range in scintillation intensity and lifetime, spectral characteristics, cost, and transparency.  It is clear to us that dual-readout is easily achievable in almost any crystal.   
 
 In all calorimeters, the energy calibration of each calorimeter volume, tower, or crystal, is simple:  for an electron beam energy of $E_0$, the mean response in scintillation light, $S_{\rm ADC}$, and the mean response in \C light, $C_{\rm ADC}$, both in units of ADC counts above pedestal, are calibrated as $E_0/S_{\rm ADC}$ and $E_0/C_{\rm ADC}$, both  constants in units of  GeV/ADC.

The simplest formulation of dual-readout response  considers hadronic showers to be composed of an electromagnetic ($e$) part and a non-electromagnetic ($h$)  part.    The $e$ part is calibrated to  unit response, and the resulting  average response of the $h$ part is denoted by $(h/e)$, which is less then unity for almost all reasonable calorimeters.  If $f_{EM}$ is the ``electromagnetic fraction'' and $E$ is the hadronic energy (either single hadron or jet), and   since the scintillation signal sums all charged particles of the shower, and the \C  signal sums predominantly only the $e$ part, the responses expected in the two signals are
\begin{displaymath}
 S= E~ [ f_{EM} + (1 - f_{EM}) ~(h/e)_S ]        
  \end{displaymath}
\begin{displaymath}
  C = E~ [ f_{EM} + (1 - f_{EM}) ~(h/e)_C ] 
\end{displaymath}
For the original {\sc dream} module, $(h/e)_S \approx 0.7$ and $(h/e)_C \approx 0.2 $\cite{h}, and the electromagnetic fraction, $f_{EM}$  event-by-event, is approximately given by $f_{EM} \approx C/E$.

\section{Fiber dual-readout calorimeters}  

The conceptual beginning of dual-readout was a talk at the Tucson calorimeter conference in 1997 \cite{tucson}, and the first dual-readout calorimeter was tested at CERN \cite{access} and intended for the International Space Station before the Challenger accident.   The first high energy physics proof-of-principle test module was the {\sc dream} module \cite{mu,e,h,em-shape,sc-separ,n1,h-shape,n2}.

\begin{figure}[hb!]
 \includegraphics[width=0.55\columnwidth]{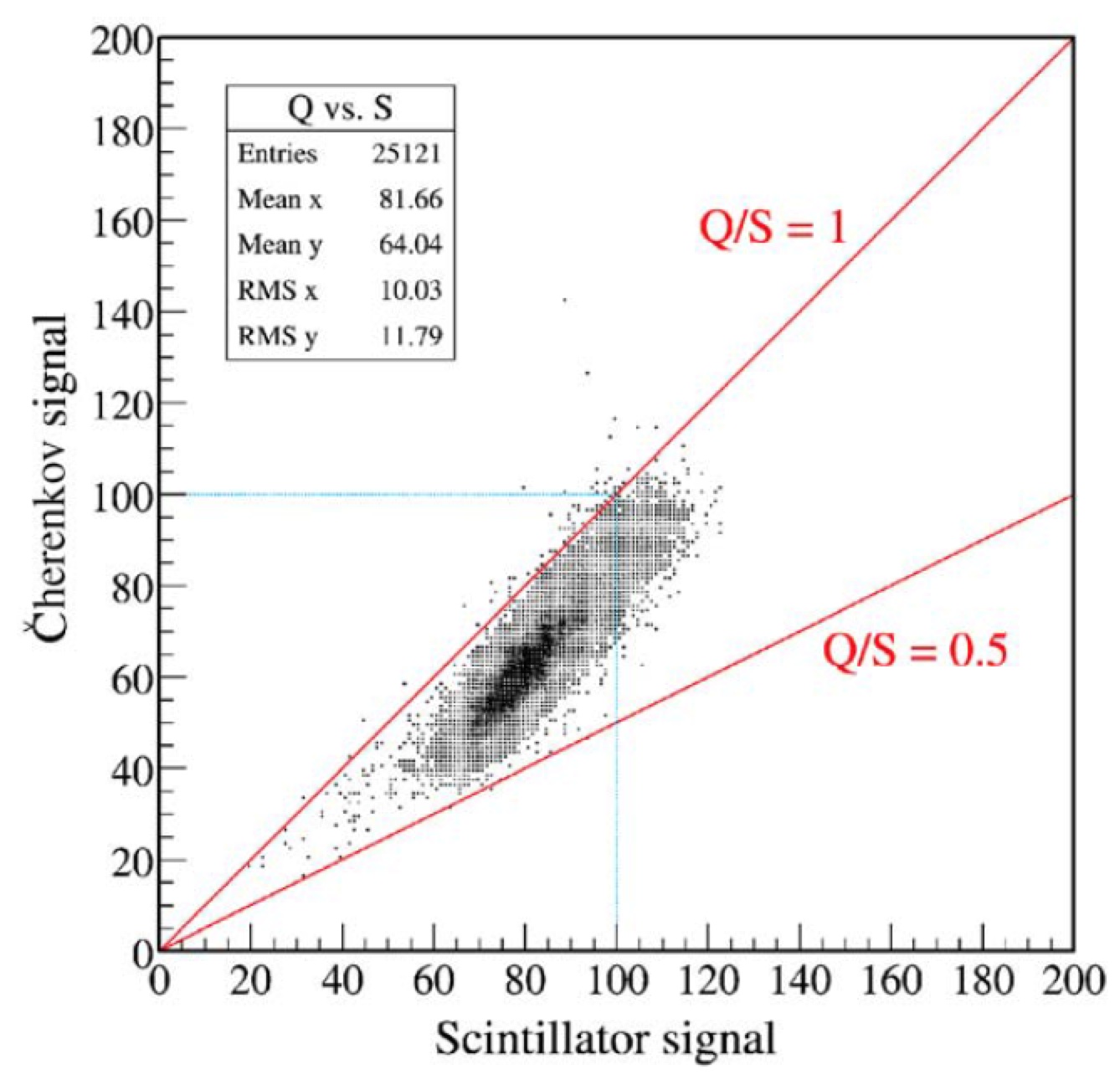}
  \includegraphics[width=0.50\columnwidth]{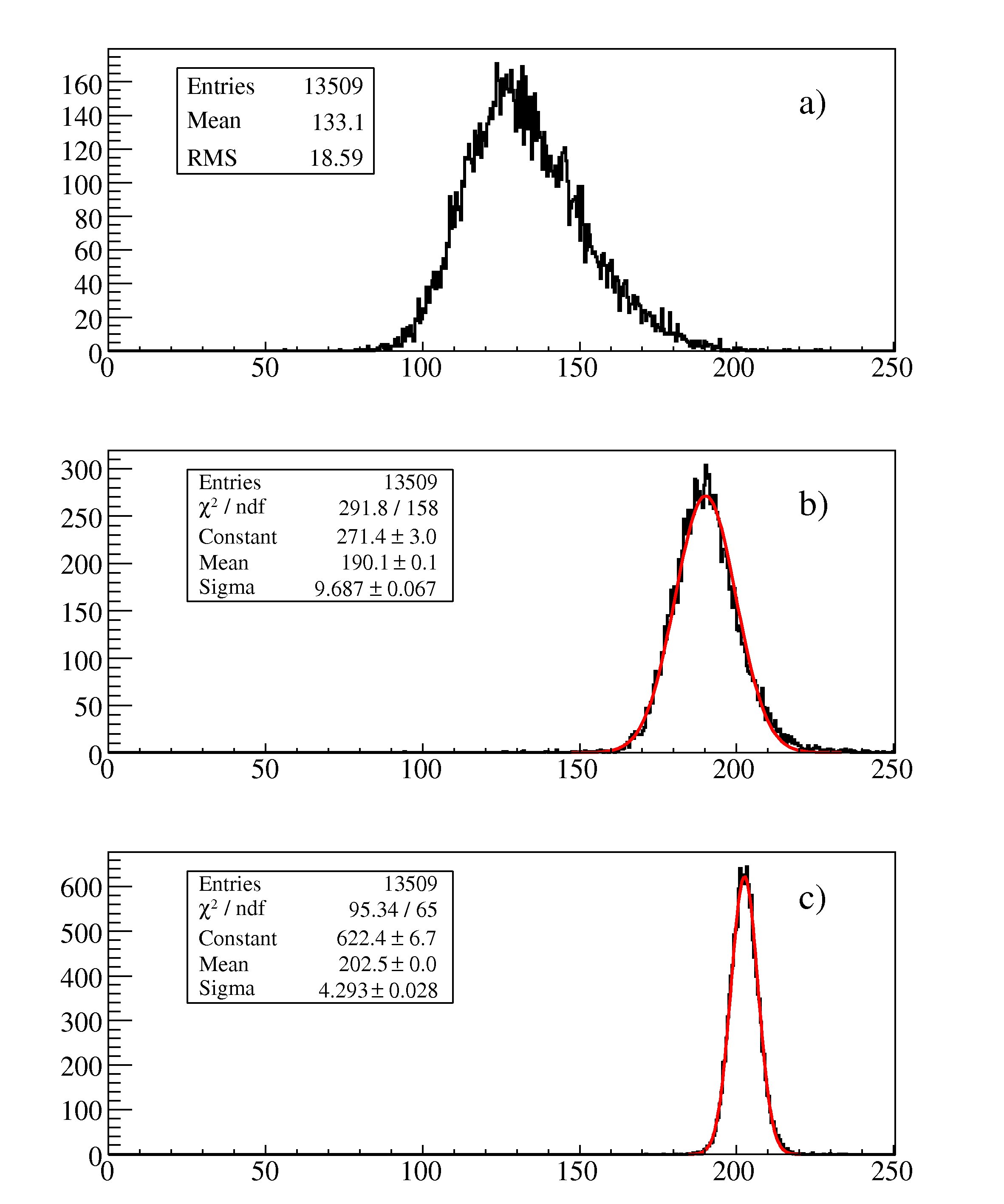}
 \caption{The characteristic ``banana'' shape of the S and C signals at 100 GeV $\pi^-$;   (a) the distribution of the S signal; (b) the distribution of the shower energy E as the solution to the two equations above; and,  (c) the solution for E when the leakage is artificially suppressed by replacing $f_{EM} \sim C/E$ by $f_{EM} \sim C/E_{\rm beam}$ for the estimate of the electromagnetic fraction.}
 \label{fig:Eres-3x}
\end{figure}

The distribution of S {\it vs.} C for 100 GeV $\pi^-$ is shown in Fig. \ref{fig:Eres-3x} in which S is always greater than C.  The distribution of S  for 200 GeV $\pi^-$ is shown in Fig. \ref{fig:Eres-3x}(a), displaying the expected energy resolution for a sampling scintillation calorimeter.  Note that (1) the mean energy is wrong; (2) the distribution is asymmetric with a high-side tail;  and, (3) the distribution is non-Gaussian.   When the above equations are solved for $f_{EM}$ and $E$, the distribution of $E$ is shown in Fig. \ref{fig:Eres-3x}(b) to have the correct energy, to be symmetric, and to be Gaussian.  These are the essential attributes of dual-readout calorimetry that make it so powerful and useful for physics.  When the beam energy is substituted for the shower energy and the electromagnetic fraction re-estimated, $f_{EM} \sim C/E \rightarrow f_{EM} \sim C/E_{\rm beam}$, the resulting distribution of $E$ is shown in Fig. \ref{fig:Eres-3x}(c).  This figure is important.  Its width corresponds to a hadronic energy resolution of 
\begin{displaymath}
  \sigma/E \approx \frac{20\%}{\sqrt{E}},   ~~ ({\rm experimentally ~expected~ resolution~ of ~ a~ {\sc dream-}like~ calorimeter})
\end{displaymath}
and this finite width is determined by the measured fluctuations in $S$ and $C$.  Therefore, a conclusion can be made that a {\sc dream}-like module (Cu absorber with scintillating and clear fibers) is capable, in the extreme case of zero leakage, of achieving the spectacular energy resolution of $20\%/\sqrt{E}$.  This resolution is close to that expected from the imperfect correlation between the unmeasurable binding energy losses in broken-up nuclei and the corresponding  neutrons that are released into the calorimeter volume \cite{wigmans},
\begin{displaymath}
  \sigma/E \approx \frac{15\%}{\sqrt{E}},   ~~~~~~~~~~~ ({\rm theoretically~ expected~ lower~ limit})
\end{displaymath}



Getting the correct hadronic energy is a big deal.  Most big detector collaborations have ``jet energy scale'' groups working continuously to establish not only the correct energies of jets, but also to estimate the systematic uncertainty on this scale.  In many physics problems at the Tevatron, for example, the uncertainly in the jet energy scale would  dominate all other uncertainties in a physics measurement.  In a dual-readout calorimeter, one calibration is done with elections of any energy into each tower and the calibration constant is known, and the hadronic energy scale is determined and correct.  In practice in the beam at CERN, we calibrate once at the beginning and once at the end of a week of data taking.

\begin{wrapfigure}{r}{0.55\columnwidth} 
\centerline{\includegraphics[width=0.50\columnwidth,height=6.cm]{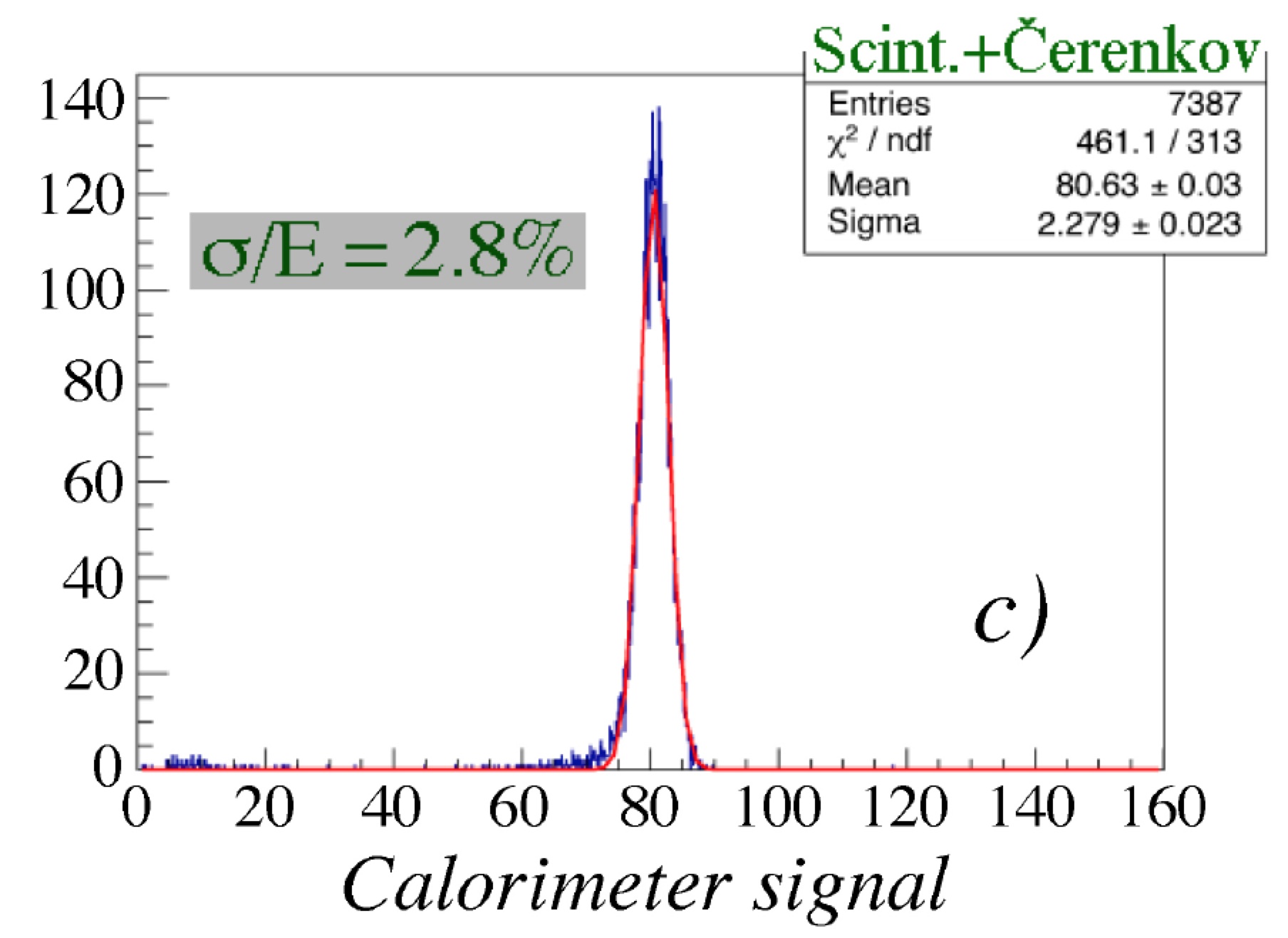}}
\caption{\dream electromagnetic energy resolution.\cite{rd52-em}}
  \label{fig:EM-resol}
\end{wrapfigure}

\begin{figure}[t!]
\includegraphics[width=5.6in]{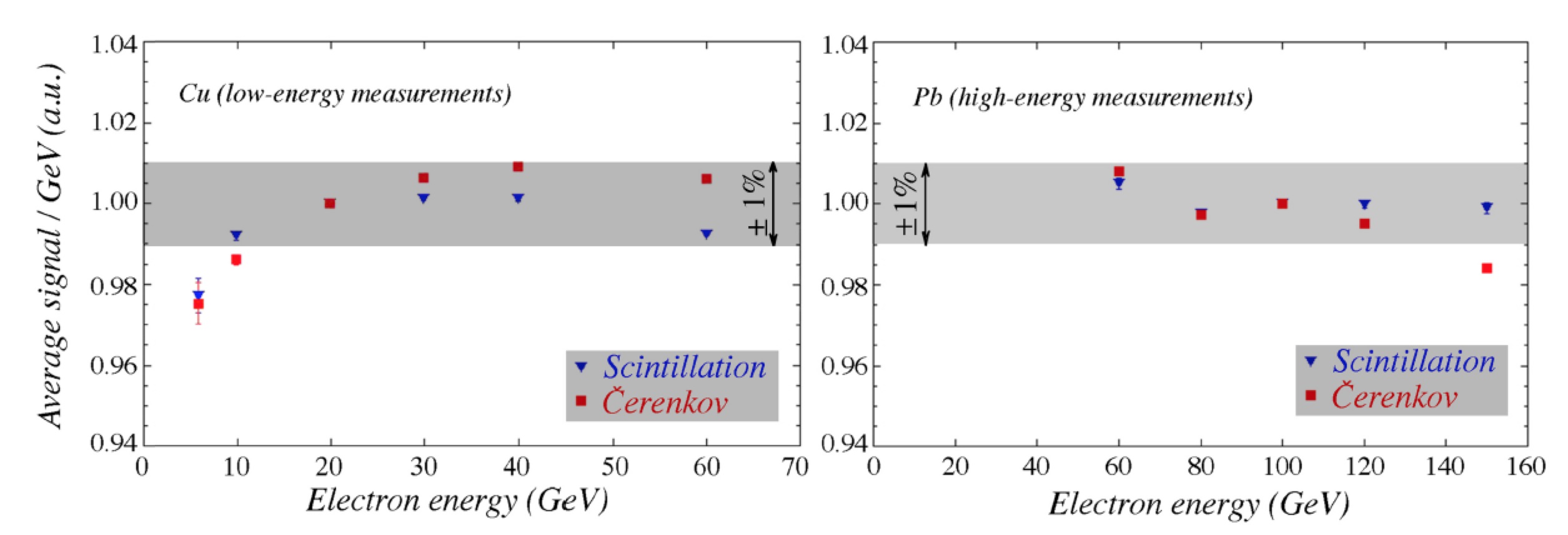}
\caption{Measured response of the dual readout calorimeters of RD52 for
  hadrons from 5 to 150 GeV.  Each module was calibrated
  only with 60 GeV electrons.  The 2\% drop at low beam energy is due to material in the beam chambers  upstream.}\label{fig:linearity}
\end{figure}

 The linearity of the original {\sc dream} module from 20 to 300 GeV is published \cite{h}, and the recently measured linearity of the new RD52 modules, both Pb-absorber and Cu-absorber, is shown in Fig. \ref{fig:linearity}.
It deserves to be emphasized that the mean values in these plots are from fits to Gaussians in each case.

\subsection{EM and hadronic \\ responses}  

The electromagnetic response of the new RD52 modules is shown in Fig. \ref{fig:EM-resol}.  This response is also Gaussian.  When the full complement of Pb and Cu modules is complete, amounting to about 5 tons of absorber with an  expected mean  lateral leakage of 1\%, we will test this calorimeter module for hadronic energy resolution.

\begin{wrapfigure}{r}{0.65\columnwidth} 
\centerline{\includegraphics[width=0.55\columnwidth]{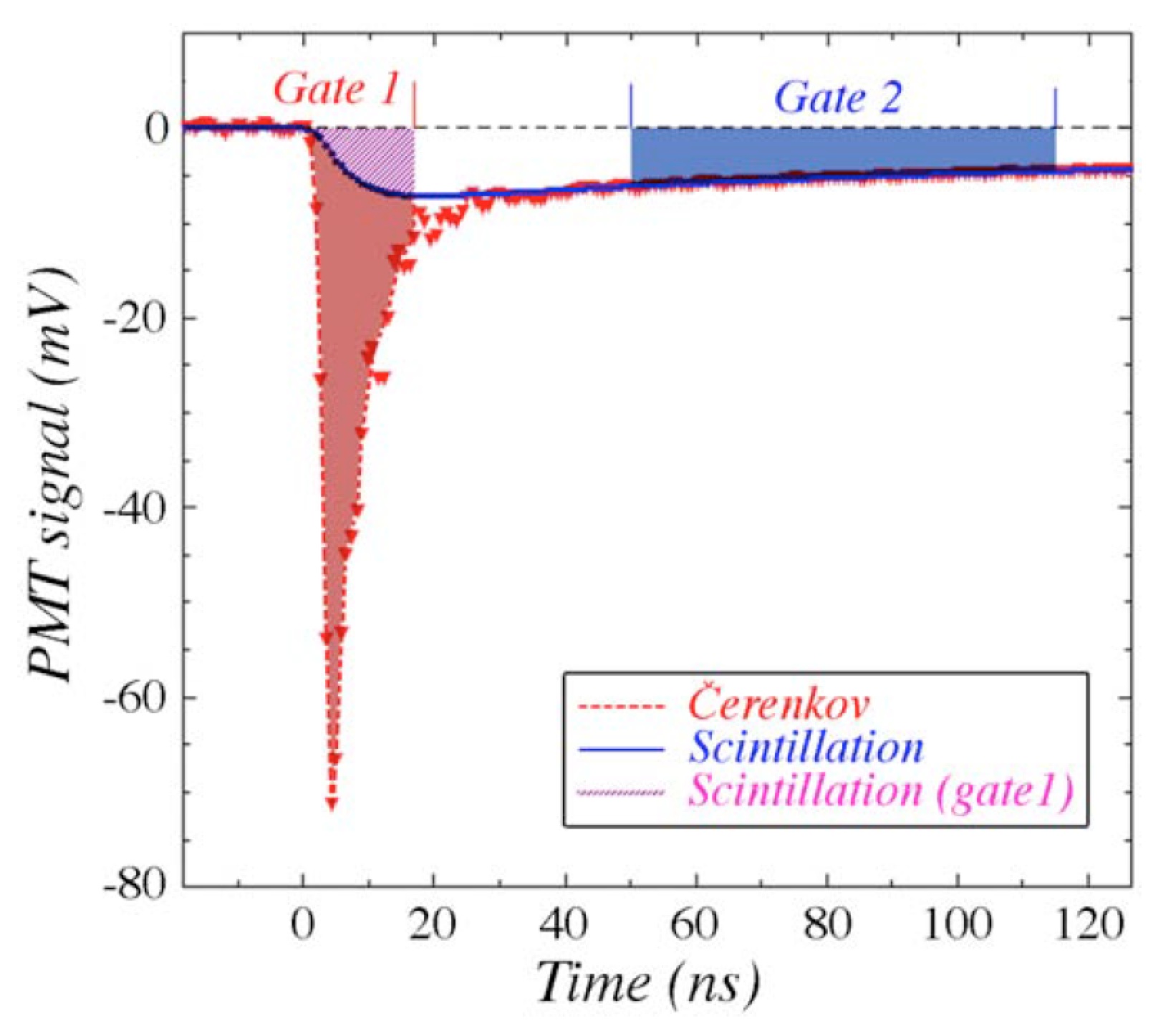}}
\caption{The PMT pulse from a BGO crystal exposed to an electron beam.  The prompt light at $t=0$ is \C light and the longer tail is the scintillation light with its expected 300ns lifetime.  Two gates on this single PMT suffice to give good measurements of the the \C and scintillation light populations.}
  \label{fig:PMT-BGO}
\end{wrapfigure}

\section{Crystal dual-readout \\ calorimeters}  

\begin{figure}[t!]
\includegraphics[width=5.6in]{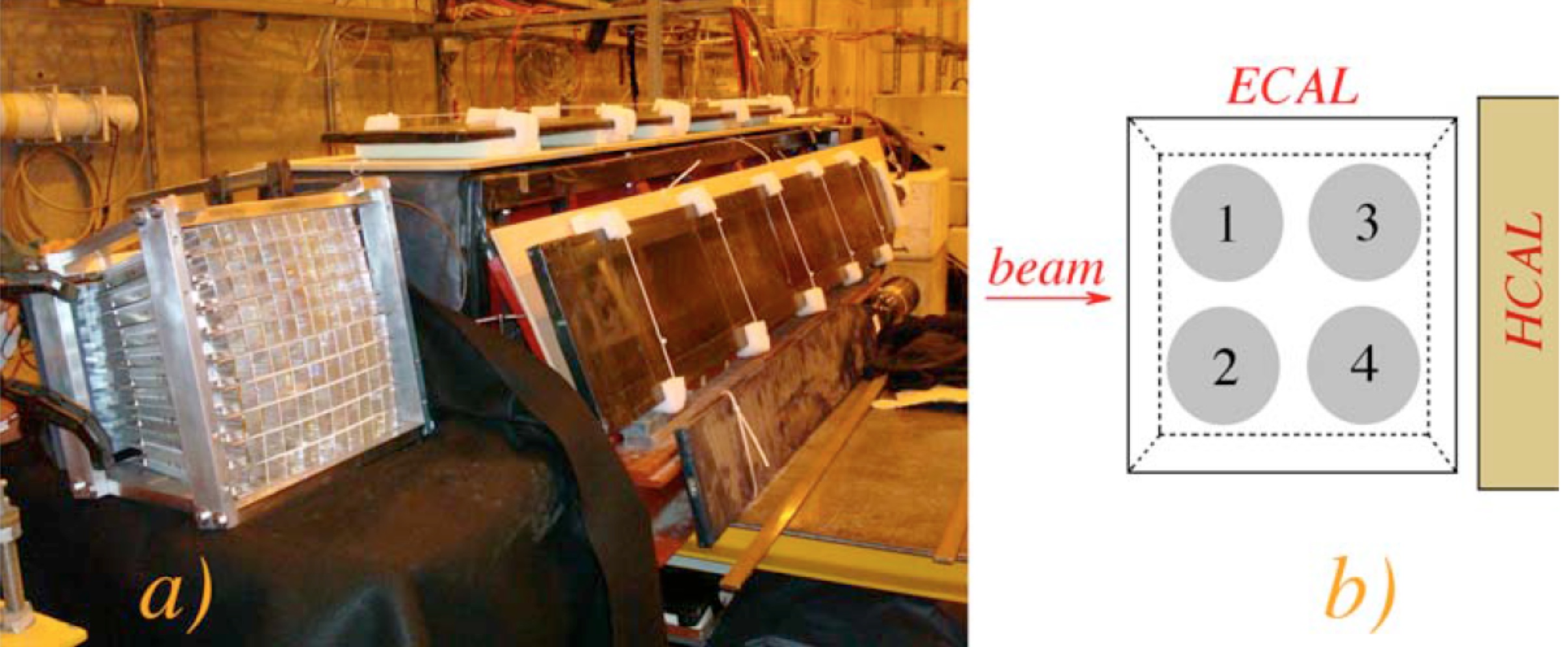}
 \caption{The 100-crystal BGO array placed in front of the {\sc dream} module and exposed to electron and pion beams incident from the lower left in this photo.}
  \label{fig:BGO+DREAM}
\end{figure}

\begin{figure}[ht!]
  \includegraphics[width=0.6\columnwidth]{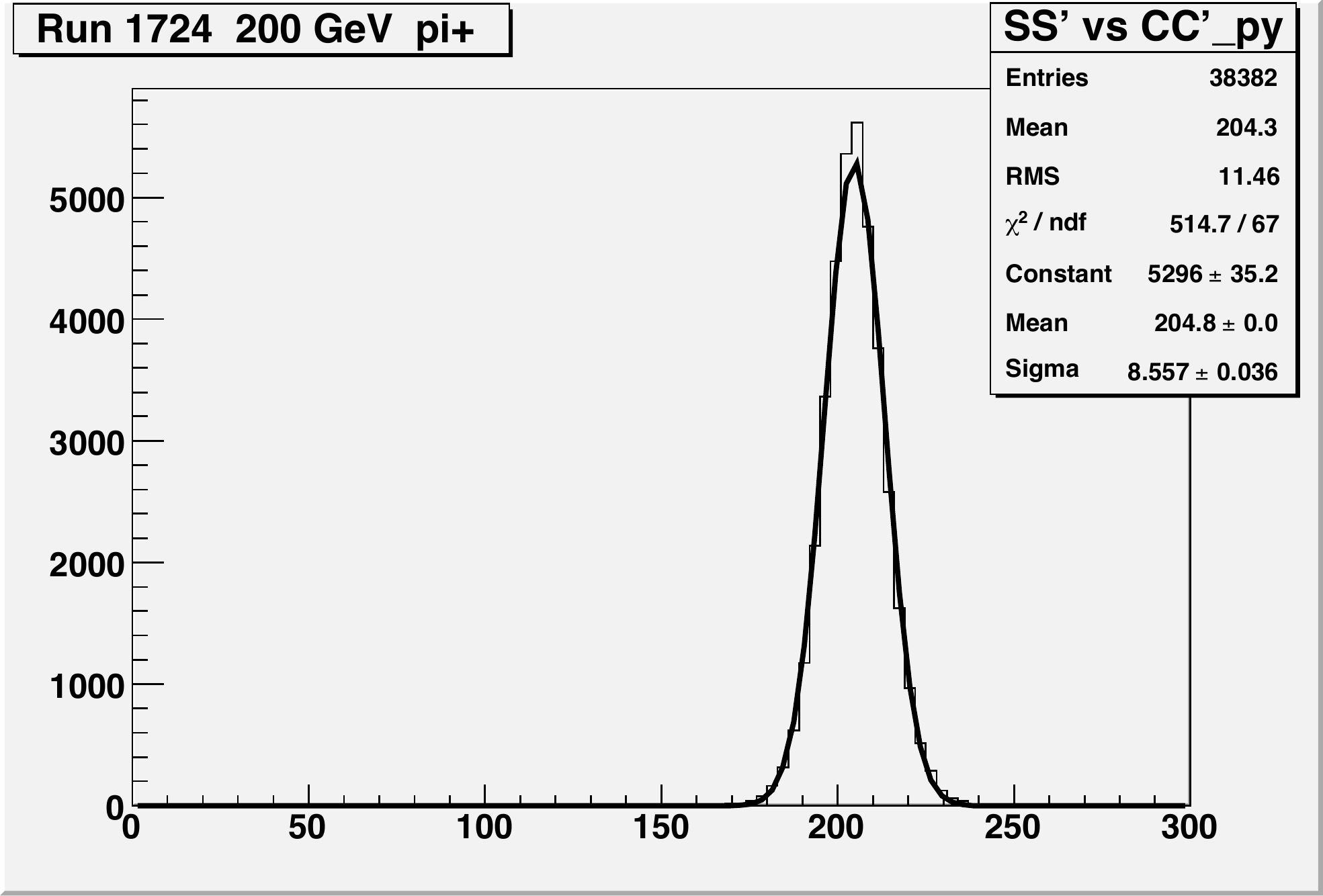}
  \includegraphics[width=0.45\columnwidth]{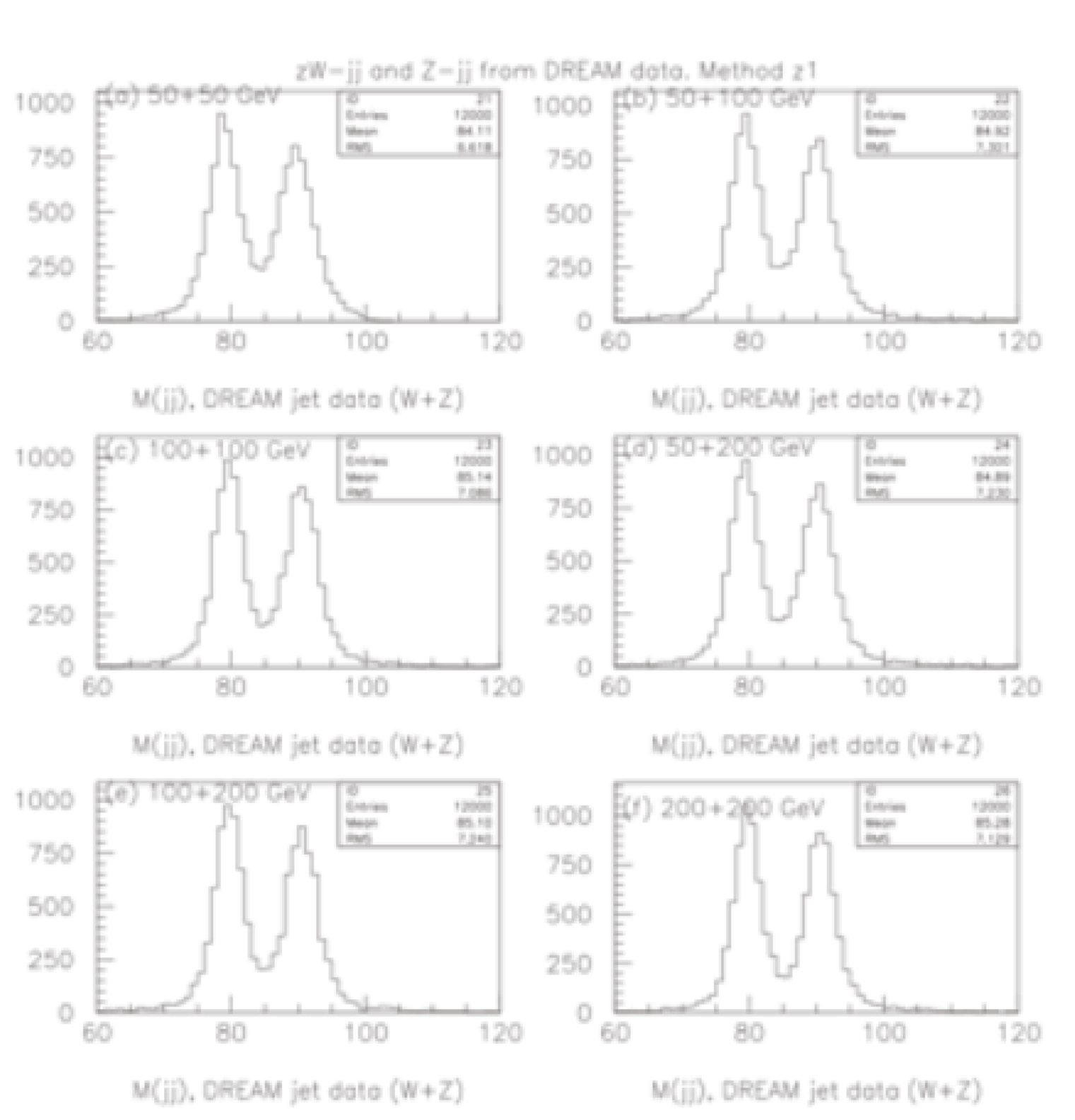}
  . \hspace{3.cm} (a) 200 GeV $\pi^+$ beam  \hspace{3.cm} (b)  $W,Z \rightarrow$ {\sc dream} jets  
  \caption{(a) Energy resolution of  200 GeV $\pi^+$ in the detector configuration of Fig. \ref{fig:BGO+DREAM}; (b) simulation by {\sc dream} data of $W \rightarrow$ jet + jet, and $Z \rightarrow $ jet + jet. \cite{DOD}}
  \label{fig:pi-200GeV}
\end{figure}

Extensive work on several different crystals is published in {\it Nucl. Instru. Meths.} 
\cite{pwo-1,pwo-2,Temp,cry-sc,pwo-Pr,crys,new-crys,bgo-em,opt-crys,pol,bgo-bso,em-crys}.  A typical separation of \C light from scintillation light in a BGO crystal is shown in Fig. \ref{fig:PMT-BGO}.  These BGO crystals were arranged in an 10-by-10 crystal array and placed in front of the {\sc dream} module, as shown in Fig. \ref{fig:BGO+DREAM}
This configuration is similar to  separate electromagnetic and hadronic dual-readout calorimeters in a collider experiment \cite{4th} and constitutes a particularly challenging combination of two very different dual-readout calorimeters.  Nevertheless, when exposed to 200 GeV $\pi^+$ beam, the hadronic energy resolution was found to be about 4\%, consistent with the expected lateral leakage from this combined BGO-crystal plus fiber calorimeter, shown in Fig. \ref{fig:BGO+DREAM}.  The energy distribution is shown in Fig. \ref{fig:pi-200GeV}(a), a nearly perfect Gaussian at the correct energy. 

\begin{wrapfigure}{r}{0.6\columnwidth}
  \includegraphics[width=7.cm,height=5.cm]{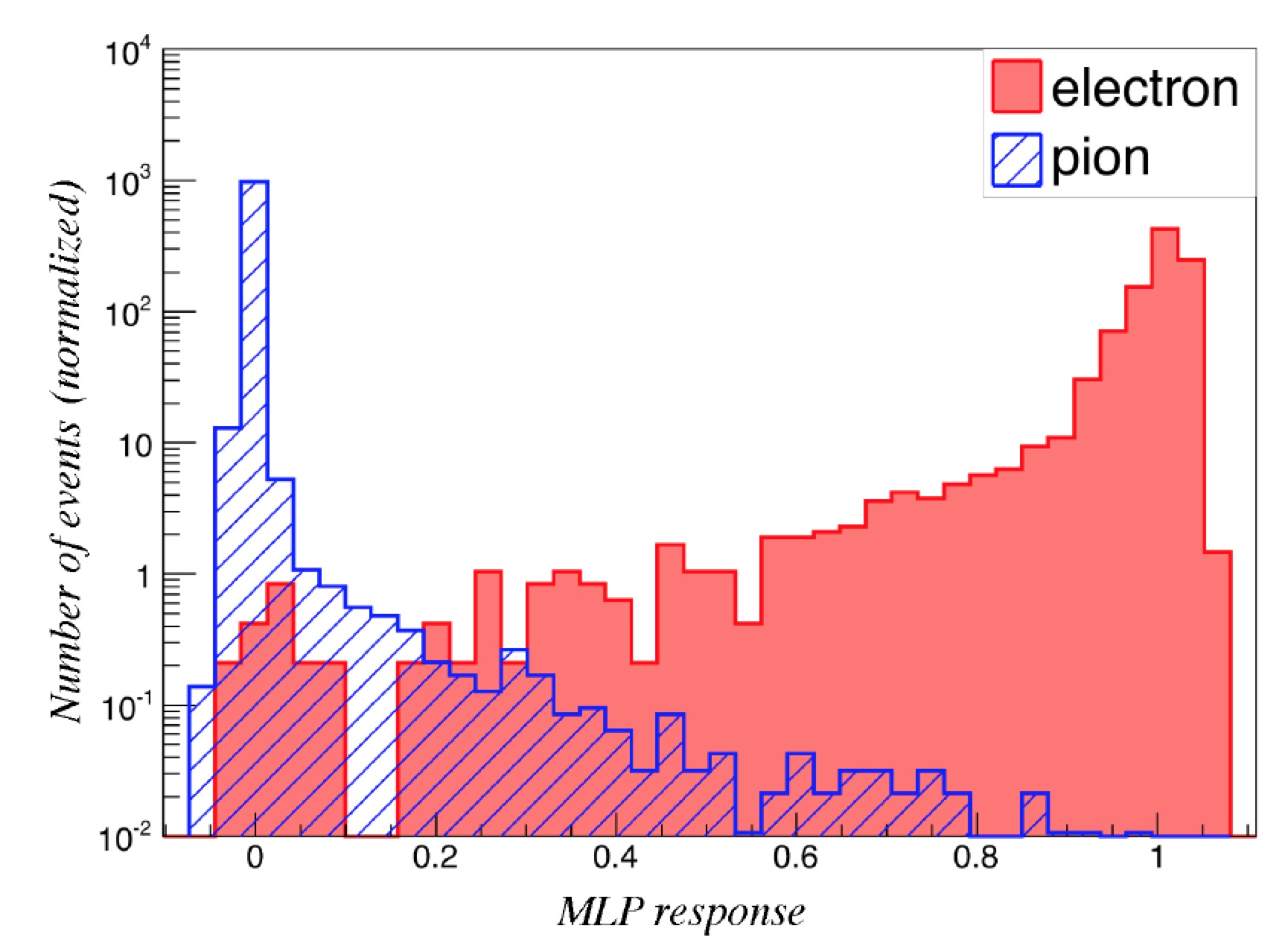}
  \caption{A multi-variate discriminant, MLP, of $e^{\pm}$ from $\pi^{\pm}$ using the fraction 
  of Fig. \ref{fig:e-pi}, an S/C measurement, and a time-depth measurement.  $\pi^{\pm}$ rejections of 99.9\% for $e^{\pm}$ efficiencies of 99.9\% are achievable.}
  \label{fig:MLP}
\end{wrapfigure}

The hadronic decays of the $W$ and $Z$ are of paramount importance in all future experiments at any collider, lepton or hadron, and measurements of the decays $W \rightarrow q^{'}  \bar{q}$ and $Z \rightarrow q \bar{q}$ to high precision are essential.  We have used {\sc dream} data to simulate these decays by randomly selecting events for the beam data files and setting the angle between the two {\sc dream} modules to properly sample the $W$ or $Z$ Breit-Wigner, and find that a dual-readout module can achieve better than a Rayleigh criterion on $W$-$Z$ separation by direct di-jet mass measurement, Fig. \ref{fig:pi-200GeV}(b).

\section{Particle Identification}  

Cogent physics in a collider experiment derives from precision measurements combined with high-probability particle identification.  In dual-readout calorimeters, we have achieved both of these in the same instrument.  

\begin{wrapfigure}{r}{0.6\columnwidth}
  \includegraphics[width=7.5cm,height=9.cm]{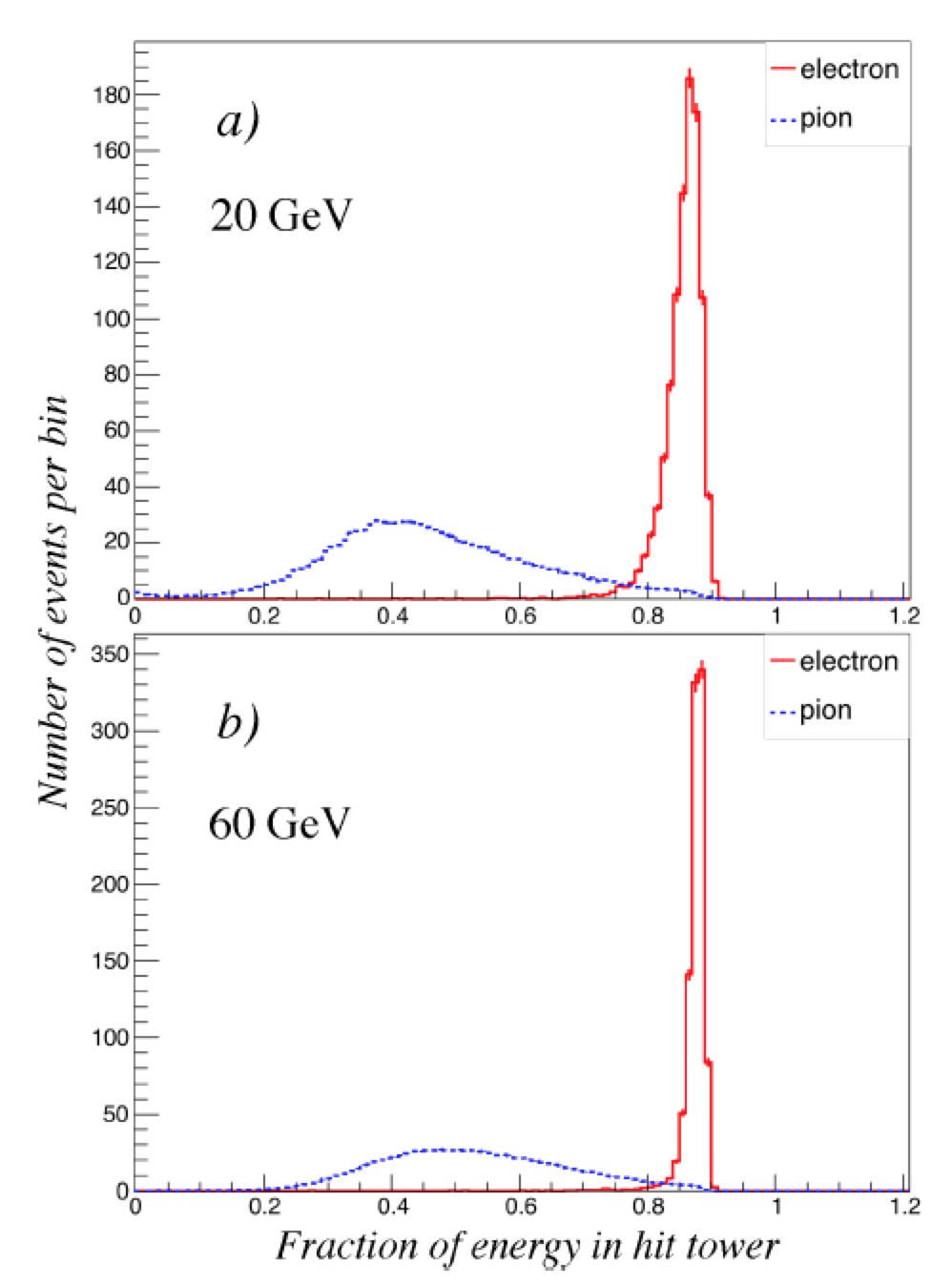}
  \caption{(a) The fraction of energy deposited in the hit tower for 20 GeV $\pi^-$'s and $e^-$'s;  (b) same for 60 GeV $\pi^-$'s and $e^-$'s.    }
  \label{fig:e-pi}
\end{wrapfigure}

A description of particle identification in all the detectors  in a collider detector is given in Ref. \cite{4th-pID}.  Here we describe only a couple of those which depend exclusively on the dual-readout calorimeter \cite{RD52-pID}, specifically $e^{\pm}$ separation from $\pi^{\pm}$ and jets, and singe $\mu^{\pm}$ separation from single $\pi^{\pm}$.  Although these are based on beam data with, necessarily, one particle at a time, these same methods and measurements are extendable to the interiors of jets with the fine lateral segmentation that is possible in a fiber dual-readout calorimeter.

The physics selected by nearly pure $e^{\pm}$ and $\mu^{\pm}$ samples includes $W^{\pm} \rightarrow \ell \nu$, $Z^0 \rightarrow \ell \bar{\ell}$, $\tau^{\pm} \rightarrow \ell \nu \bar{\nu}$, and $H^0 \rightarrow W^+ W^-$, $H^0 \rightarrow e^+ e^-$, $H^0 \rightarrow \mu^+ \mu^-$, $H^0 \rightarrow \tau^+ \tau^-$ decays, and others with tertiary decays to leptons.  Of course, lepton identification is also a window to unknown physics.

 \subsection{$e^{\pm}$ separation from $\pi^{\pm}$ and jets $(j)$}
 
 Electron showers are very narrow and one criterion is the fraction of a particle's energy that is contained in the calorimeter tower this is hit.  These distributions are shown in Fig. \ref{fig:e-pi} for electrons and pions at two energies, 20 GeV and 60 GeV.  By eye, the separation is close to 100-to-1.  Combining this measurement with the timing, similar to that achieved in {\sc spacal} \cite{spacal}, and also with an $S/C$ measurement, the combined separation is shown in Fig. \ref{fig:MLP} after a multi-variate analysis \cite{RD52-pID}.  The rejection here is about 1000-to-1 while retained an exceedingly high electron acceptance of 99.9\%.  These numbers are so favorable that, in a real $4\pi$ experiment with backgrounds of many sorts, the final rejection-efficiency relation may not be so favorable, but this is a good starting point. 
\begin{wrapfigure}{r}{0.6\columnwidth}
\centerline{\includegraphics[width=0.55\columnwidth]{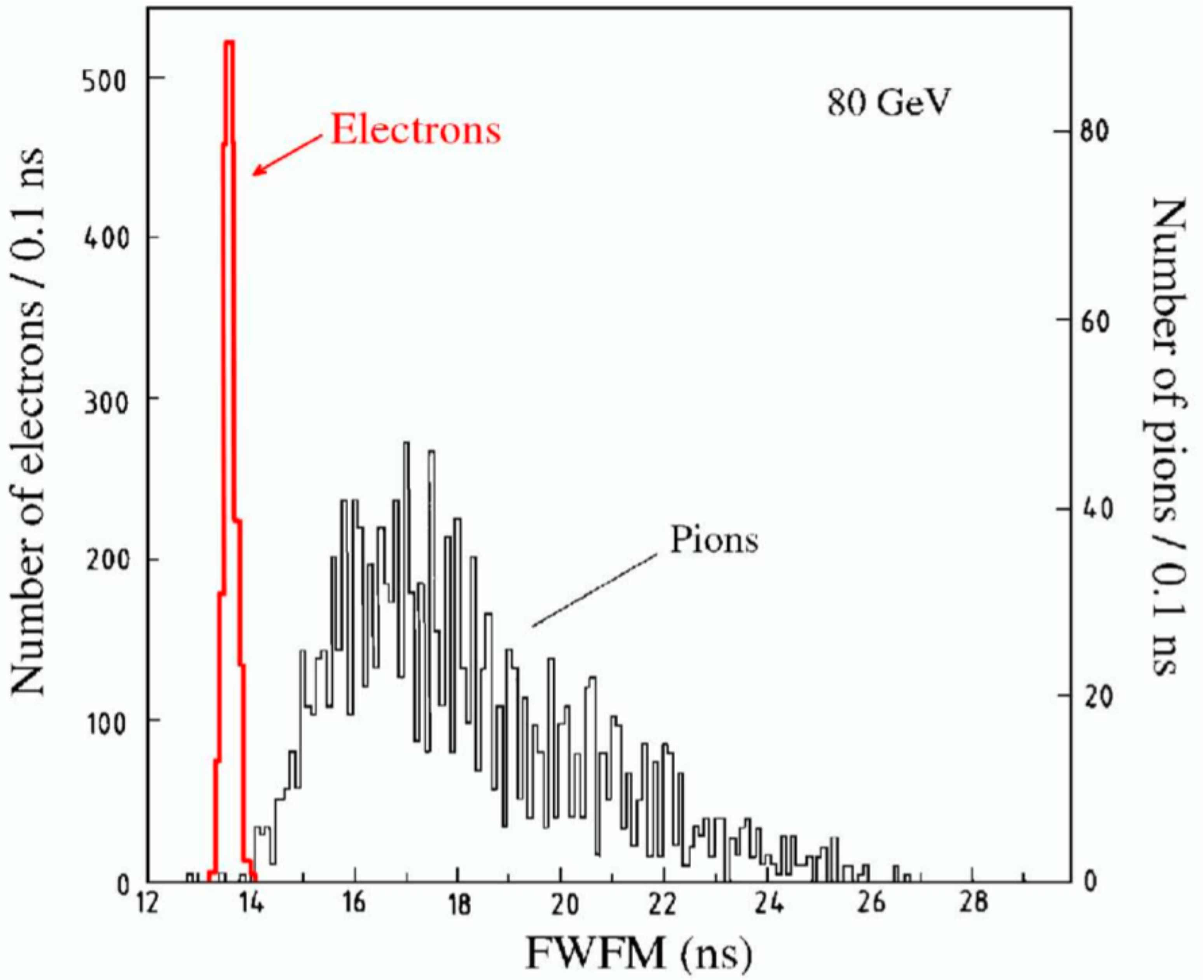}}
\caption{The full-width at one-fifth maximum of the scintillation signal in {\sc spacal}.}\label{fig:spacal}
\end{wrapfigure}
A further independent measure was made by {\sc spacal} \cite{spacal} and shown in Fig. \ref{fig:spacal} for 80 GeV $e^-$ and $\pi^-$.    Simply put, all electron showers look the same in time and space, whereas pion initiated showers contain large fluctuations in electromagnetic fraction, neutron content, and spallation proton energy deposits.  These appear as a large variations in the time is takes for the last of these particles to come to rest.  This time is represented as the ``full width at one-fifth maximum (FWFM)'' on the $t$-axis.  Evidently, the separation is about 100-to-1.  This time measurement is similar to one of the three measurements that were included in the MLP discriminator \cite{RD52-pID}.    
 
Clearly, this same discriminant can be implemented in a dual-readout calorimeter using only the scintillating fibers and, more importantly, we expect to read out all channels at 5 GHz to have complete control over the  optimization of  this discriminator.  Also, it is the long-time tail of this scintillation signal that yields the neutron signal in the {\sc dream} and RD52 calorimeters.

\subsection{$\mu^{\pm}$ separation from $\pi^{\pm}$}

\begin{wrapfigure}{r}{0.6\columnwidth}
\centerline{\includegraphics[width=0.55\columnwidth]{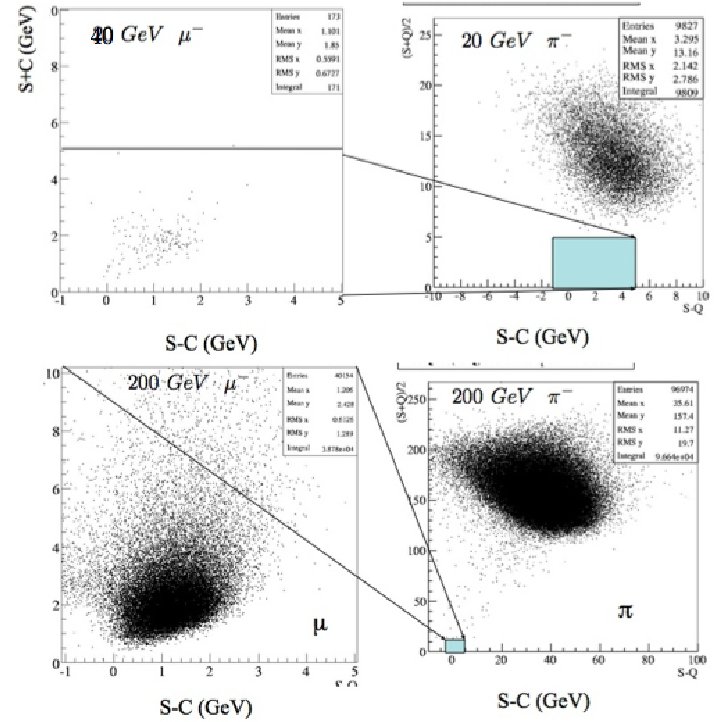}}
\caption{Dual-readout separation of $\mu-\pi^{\pm}$.}\label{fig:mu-pi}
\end{wrapfigure}

$\mu^{\pm}$ that traverse the small 1-ton {\sc dream} module approximately aligned with the fibers allow a unique $\mu^{\pm}$ identification:  the \C angle is larger than the capture angle of the clear fibers and, therefore, the  \C photoelectron signal is zero.  The scintillation signal is, as expected, about 1.1 GeV for the 2-meter module.  In the event of a hard bremsstrahlung or pair production by the muon within the body of the {\sc dream module}, that energy is purely electromagnetic for which the scintillation and \C responses are equal.   Therefore, for a muon, $S-C \approx 1.1$ GeV, but it is  different for a $\pi^{\pm}$.  Distributions   of $(S-C) ~vs. ~ (S+C)/2$ for $\mu^-$ in the {\sc dream} module \cite{mu} are shown in Fig. \ref{fig:mu-pi} for 20 GeV $\mu^-$ and $\pi^-$ and 200 GeV $\mu^-$ and $\pi^-$.   Clearly, the discrimination is excellent at higher energies, but it is still as good as 1000-to-1 at 20 GeV.  We have not been able to test to lower beam energies in the H2 beam at CERN, but will resume this work in the H8 beam at LHC start-up.

Of course, the calorimeter absorber itself provides for muon-pion discrimination and, in the 4th detector \cite{4th} the second magnetic field region in the annulus between the dual-solenoids, even further muon-pion discrimination is possible with an energy balance criterion using the main tracker, the calorimeter, and the second muon momentum measurement.  Rejections as large as $10^5$-to-$1$ were calculated, but clearly confusion and other effects will limit this particle identification long before $10^5$ is reached.  Nevertheless, this is a good starting point for $\mu^{\pm}$ identification.

Dual-readout is rich in unique and powerful particle identification measurements from which we have only discussed a few.  In fact, it was argued in \cite{4th} that all patrons of the standard model are identifiable with a sufficiently well designed calorimeter inside a good detector.

\section{Summary}

We understand the fiber and  crystal dual-readout modules that we have built, and the performance of these modules is limited mainly by lateral leakage.  To solve this, we are building a large module that is expected to have a mean energy leakage of about 1\%, and this large calorimeter consists of separate smaller Pb-absorber and Cu-absorber modules. 

Our near-term work in the next two years consists of (i) the building of  a one-ton Cu dual-readout module by rolling the Cu, (ii) the building of a small W-absorber module, (iii) the completion of a 5-ton module that will function both as a Pb-module and a Cu-module, and (iv)  the direct comparison of  the fundamental limitations to hadronic calorimetry in Pb, Cu, and W.  In parallel, we will (v) devise and test new optical readout schemes other than PMTs, (vi)  formulate a {\sc geant4}-based simulation of RD52 measurements, (vii)  design and test a gaseous dual-readout module, and (viii) design fiber wedges for a $4\pi$ calorimeter.

The two reports to the SPS Council \cite{spsc-12,spsc-13} are a more complete and more technical description of the present and future expectations of RD52.




\begin{footnotesize}

\end{footnotesize}


\end{document}